\documentclass[epj,referee]{svjour}
\usepackage{multirow}
\usepackage{graphicx}
\begin{document}

\title{Rearrangement collisions between gold clusters\thanks
  {Supported by the \textit{Fondo Nacional de Investigaciones
      Cient{\'\i}ficas y Tecnol{\'o}gicas} (FONDECYT, Chile) under
    grant \#8990005 and \#1010988 and DIPUC-Chile.}  }

\author{Jos{\'e} Rogan \inst{1} \and Ricardo Ram{\'\i}rez \inst{2} \and
  Aldo H. Romero \inst{2}  \and Miguel Kiwi \inst{2}
}                     
\offprints{Miguel Kiwi}          
\institute{Departamento de F{\'\i}sica, Facultad de Ciencias,
Universidad de Chile, Casilla 653, Santiago 1, CHILE \and Facultad de
F{\'\i}sica, Universidad Cat{\'o}lica de Chile, Casilla 306, Santiago,
CHILE 6904411} 

\date{Received: date / Revised version: date}
%

\abstract{ Collision processes between two gold clusters are
  investigated using classical molecular dynamics in combination with
  embedded atom (EA) potentials, after checking the reliability of EA
  results  
  by contrasting them with first principles calculations. The Au
  projectiles considered are both single atoms (N=1) and clusters of
  N=2, 12, 13 and 14 atoms. The targets contain N= 12, 13 and 14
  gold atoms. The initial projectile energy $E$ is in the range $0 <
  E < 1.5$~eV/atom. 
  The results of the collision processes are described and analyzed in
  detail.
} 
\maketitle

\section{Introduction}
\label{intro}

The study of nanostructures has recently attracted wide\-spread
interest among theoretical and experimental physicists and chemists,
and because of its many applications has also come to the forefront of
technology~\cite{rev-heer}. On the theoretical side {\it ab initio}
procedures are now capable of providing incisive insights into the
properties of these systems. In addition, novel and sophisticated
nanostructure fabrication, manipulation and measurement techniques
have given impetus to experiment, and reliability to a large amount of
experimental data~\cite{clev97,scha97,tayl92,koga98,spas98,palp98}.
Moreover, the technological applications on a 
variety of devices has strongly stimulated activity in the
nanostructure field since they can be considered as building blocks
of novel nanostructured materials and
devices~\cite{whet96,andr96,mirk96,aliv96}.

In particular, metallic clusters provide an interesting subject of
study for at least two reasons: i)~clusters constitute intermediate
systems between isolated atoms and molecules, on the one extreme, and
bulk solids on the other ({\it i.e.}  they constitute genuine
mesoscopic systems); and ii)~often they exhibit an interesting
phenomenology of their own. Gold clusters have received widespread
attention during the last two decades, both
experimentally~\cite{clev97,scha97,tayl92,koga98,spas98,palp98} and
theoretically~\cite{zhao94,hand94,garz96,habe97,wang01,garz98,belb98,barn99,sole00,wils00,liyi00,hakk00,gron00,rome02}.

In principle one does expect {\it ab initio} procedures to be the
definitive tool to handle this type of systems; however, it is not
always feasible (or at least practical) to implement {\it ab initio}
calculations. In fact, since we are interested in structures with a
fairly large number of atoms, or arranged in several different
interacting nanostructures, {\it ab initio} computations first become
very time consuming and, in the end, impractical. Much the same
happens when attempting to obtain a detailed description of the long
time evolution of small systems, or to describe their properties when
subject to a large variety of external conditions. 

Here we intend to develop an adequate description of colliding gold
clusters, a phenomenon which falls into one or more of the categories
described in the preceding paragraph. In this paper we employ mainly
classical molecular dynamics, an alternative computation scheme that
has proven to be quite reliable~\cite{sole00}. Classical molecular
dynamics (MD) is a valid option in this case, since it
allows to considerably reduce the computation time (relative to MD
{\it ab initio} calculations), and/or to increase substantially the
number of particles that can be handled.  Obviously, there is a price
to pay; as examples of this cost we mention that the method is at best
semi-classical, that the detailed electronic structure is ignored, and
that phenomenological potentials, adjusted to fit bulk properties, are
used. In spite of these shortcomings the MD technique can reliably be
used to compute the properties dominated by the ionic contribution,
which is the case for the phenomena treated in this paper. This is
especially true when MD is checked, in some physical limit, against
first principle calculations.  An alternative, but equally convenient
procedure, is to close in on a solution, for example for geometrical
optimization via MD, to be followed by first principle calculations.
The synergy between {\it ab initio} and MD thus allows to
significantly reduce the resources that are required and to expand the
set of problems amenable to treatment.

More than a decade ago the bombardment with gold clusters of metallic
surfaces was investigated experimentally~\cite{matt86} and
theoretically~\cite{eche90,shap90,shap91}.  Moreover, the dramatic
energy accommodation that occurs in cluster-cluster collisions, which
is crucial to understand the growth mechanism during the early stages
of particle formation, was investigated around the same time by
Blastein {\it et al.}~\cite{blas92}.  We focus our interest on
the dynamics of gold cluster collisions. In particular we study the
structure and symmetry of small Au clusters, and the dynamics of
atom--cluster and cluster--cluster collisions, for different values of
the impact parameter $b$ and as a function of center of mass energy $E$.

This paper is organized as follows: after this Introduction we
describe, in Sec.~\ref{sec:simulation}, both the {\it ab initio} and
EA methods employed in the computations. In Sec.~\ref{sec:results} we
provide results, for a large variety of cases, of the
implementation of the codes, for the cluster structures, their
symmetries and the dynamics of the collision process. Finally, in
Sec.~\ref{sec:concl} we draw conclusions and close the paper.

\section{Simulation method}
\label{sec:simulation}

\subsection{Ab-initio method}

\begin{figure}[tbp]
\begin{center}
\includegraphics[angle=0, width=12.5cm]{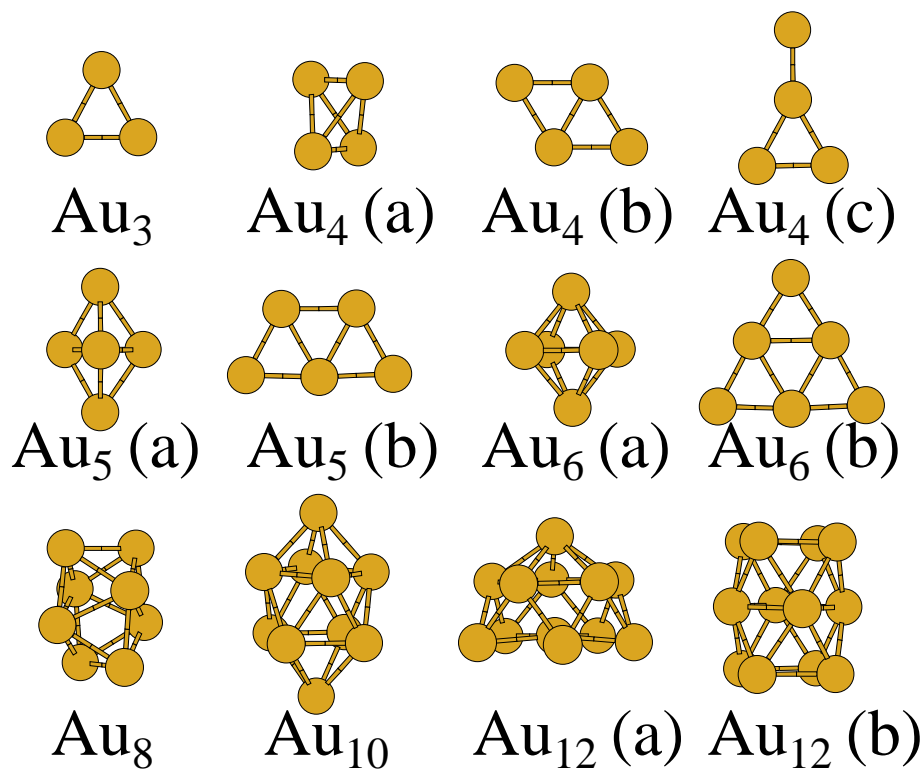}
\vskip .5cm 
\caption{\textit{Ab initio} optimized geometries for 3, 4, 6, 8, 10
  and 12 atom gold clusters.}   
\label{fig:abgeo}
\end{center}
\end{figure}

In order to asses the quality of the semi-empirical embedded atom (EA)
procedure used in the context of our classical molecular dynamics
simulations, we start contrasting EA against first principles
calculations. Thus, {\it ab initio} geometrical optimization was
carried out within the Car-Parinello approach~\cite{carpar85}, in the
framework of the density functional theory, using gradient corrections
in the PBE implementation~\cite{PBE}. Gradient corrected functionals
have been adopted in recent theoretical studies of geometrical
optimization of metallic clusters, mainly because they are more
accurate than the local density functional, even though there still is
some controversy in the literature on the ground state geometry of
small Au clusters~\cite{habe97,part90,semi97}. The calculations were
performed only at the $\Gamma$-point of the Brillouin zone, using norm
conserving pseudopotentials~\cite{trou91}. The wavefunctions were
expanded in plane waves, with an energy cutoff of 60 Ry. We have
explicitly checked that, with this energy cutoff, the structural
properties of our system are well converged. The box used in the
calculations was always at least three times larger than the cluster
diameter. The results obtained, as well as their EA counterparts, are
given in Fig.~\ref{fig:E_b} and in Table~\ref{table:binding_E}. The
{\it ab initio} optimized geometries are illustrated in
Fig.~\ref{fig:abgeo}. The agreement between {\it ab initio}, EA and
Wilson and Johnston~\cite{wils00} (WJ) procedures is quite
satisfactory for large clusters (N$\le$10) and provides a reasonable basis
to trust the EA calculations that constitute the core of the present
paper.

\onecolumn

\twocolumn

\subsection{Embedded atom method}

The interatomic interaction between gold atoms is modeled using a
semi-empirical (EA) potentials~\cite{daw84,foil86}. On the basis of
these EA potentials we obtain the average binding energy per atom,
$E_b$, which is later minimized to yield the optimal cluster geometry.
The latter is achieved using a Monte Carlo procedure, for which we
adopted as starting configurations, for the different cluster sizes,
the geometries found by Wilson and Johnston~\cite{wils00} (WJ).  Once
the optimal geometry is established several static properties, like
nearest neighbor distances and angles, and the average coordination
number, are readily evaluated.

After the various different clusters are properly characterized we use
classical molecular dynamics (MD) to simulate the cluster--cluster
scattering process. Many body EA semi-empirical potentials are used
throughout. To integrate the equations of motion we implement the
Verlet velocity algorithm, with a 1 femtosecond time step.  Since the
collision fragments heat up as a consequence of the scattering process
it is necessary to cool them down; to do so we simply rescale the
temperature by 1~\% every 1000 steps, during a total of 100~000 time
steps, to reach a final energy of 63~\% of the initial one.  Finally,
the collision fragments are carefully scrutinized to extract the
physical information we are looking for.

\section{Results and Discussion}
\label{sec:results}

\subsection{Structures}

The first issue we address is to check the reliability of our MD
procedure as compared with alternative methods. In
Table~\ref{table:binding_E} we compare the average binding energy
$E_b$ of the lowest energy configurations, and the average nearest
neighbor distance $R$, of clusters built with different numbers of Au
atoms. On the one hand we have {\it ab initio} results obtained within
the framework of density functional theory, and on the other the
empirical potential results of WJ~\cite{wils00} (who used the
Murrell--Mottram~\cite{murr90,cox99} potential) as well as the EA
values that we obtained.

\begin{figure}[tbp]
\includegraphics[angle=-90, width=8.5cm]{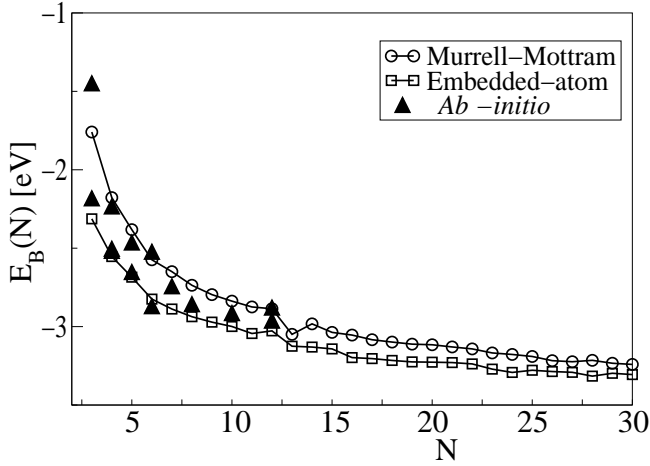}
\vskip .5cm 
\caption{Binding energies per atom $E_b$, obtained by WJ using the
Murrell--Mottram potential, and by us on the basis of the EA
potential and {\it ab initio}, as a function of the number of atoms in
the cluster N. The {\it ab initio} results, for which we have
considered several possible geometries, are detailed in
Table~\ref{table:binding_E}.}   
\label{fig:E_b}
\end{figure}

It is quite apparent that the EA estimates for $E_b$ differ 
considerably from the ones found by WJ, and also with the
more trustworthy {\it ab initio} results. However for small clusters
the geometrical parameters are not so satisfactory.  This small N
(where N is the number of atoms in the cluster) error margin is not
unexpected, since the EA potential has been adjusted to fit bulk
properties and cannot be expected to fully succeed in systems where N
is tiny.  However, and also as expected, the situation improves as N
increases, which is precisely what is borne out by
Table~\ref{table:binding_E}, where we display results for the 
range of clusters sizes from N=3 to 12.  The same trend is observed in
Fig.~\ref{fig:E_b}, where we plot $E_b$ versus N. We notice that {\it
  ab initio} calculations were performed for N = 4, 5 and 6 planar
and three dimensional structures.  It is apparent that as the cluster
size increases, specially when the number of atoms is larger than 10,
that both the {\it ab initio} and EA values tend towards the bulk value.

\begin{figure}[tbp]
\includegraphics[angle=-90, width=8.5cm]{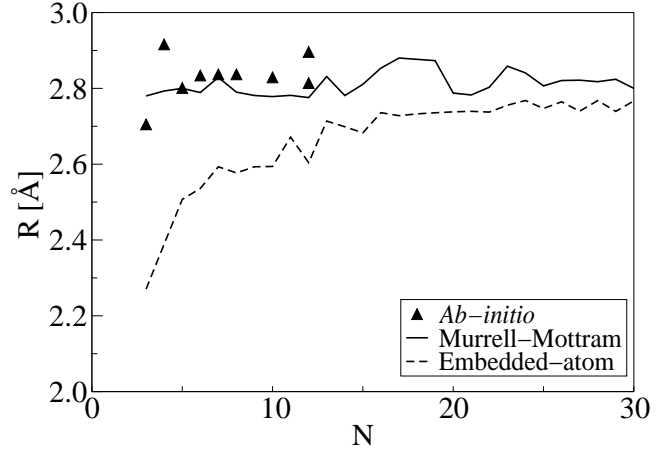}
\vskip 1.cm 
\caption{Average nearest neighbor distances R calculated using EA
  (dashed line), and those obtained by WJ using the Murrel-Mottram
  potential (full line).}
\label{fig:distancias_vs_N}
\end{figure}

The average nearest neighbor distances $R$ that we compute also
exhibit larger errors than those of WJ.  In spite of the fact that we
employ the WJ cluster configurations as the starting point for our
calculations, but in which we use a binding energy obtained from a
different potential, we derive geometrical structures which differ
from those of WJ. However, once again, increasing N yields compatible
results.  For example, for a 3-atom cluster the difference in nearest
neighbor distances amounts to 15\%, but already for a 6-atom cluster
it reduces to only 8\%. These average distances are illustrated in
Fig.~\ref{fig:distancias_vs_N}, where we plot $R$ as a function of
the number of atoms in the cluster, N. In contrast with
Fig.~\ref{fig:E_b} the plot of Fig.~\ref{fig:distancias_vs_N} is not
smooth, but shows abrupt variations between two successive values of
N. Despite this roughness the tendency of the WJ and our EA plots is
to approach each other as N$\gg 1$. Moreover, our results are in good
agreement with the {\it ab initio} ones obtained by
Wang~\cite{wang01}.

\subsection{Symmetries}

\begin{figure}[!h]
  \centering
  \includegraphics[angle=-90, width=7.5cm]{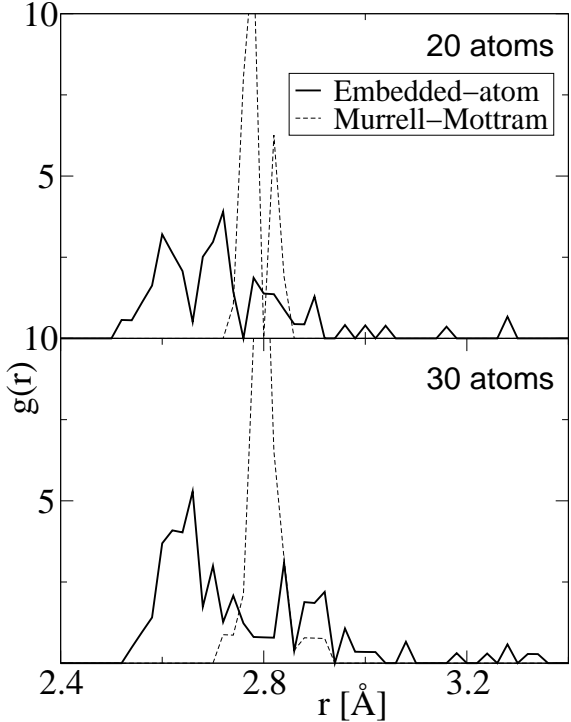}
  \caption{Pair correlation function $g(r)$ for 20, 30 and 40 Au atom
    clusters as calculated by us, using EA potentials, and by WJ.}
  \label{fig:fcp}
\end{figure}

In addition to the binding energies and interatomic distances the
cluster symmetry is a relevant characteristic and, in the context of
gold cluster topologies, the Jahn-Teller effect is also an important
element. {\it Ab initio} calculations predict a C$_{2v}$ symmetry for
Au$_3$ and Au$_4$, while EA yields C$_{2v}$ and D$_{2d}$, and the
Murrell--Mottram potential used by WJ yields D$_{3h}$ and T$_d$
symmetries, respectively. These differences are quite apparent in the
pair correlation function $g(r)$ plotted in Fig.~\ref{fig:fcp} .

\begin{figure}[!h]
\vskip 1.cm
  \centering
  \includegraphics[angle = -90, width =8.cm]{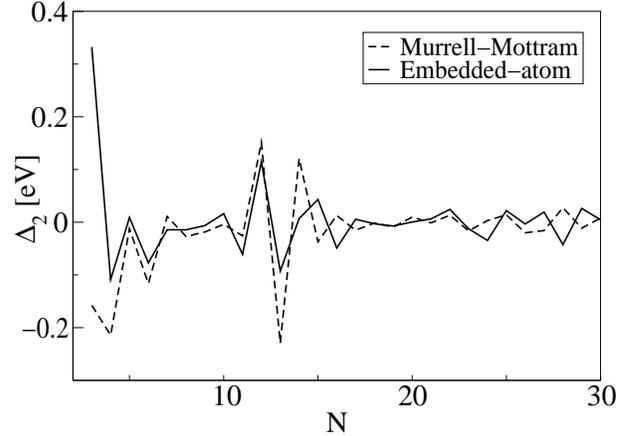}
  \vskip 5mm
  \caption{Second difference in the binding energy $\Delta_2
    \{E_b(\mbox{N}) \}$ as a function of cluster size N.}
   \label{fig:secdif}
\end{figure}

The second difference in the binding energy is defined by
\begin{equation}
\label{Delta2}
\Delta_2E_b(N)=2E_b(N)-E_b(N-1)-E_b(N+1)\ ,
\end{equation}

\noindent and gives an indication of the stability of a cluster with
respect to disproportionation~\cite{rev-heer}, as well as its ionic
hardness. A plot of $\Delta_2E_b(N)$ as a function of N is given in
Fig.~\ref{fig:secdif}, where we observe a good agreement of our EA
values with those reported by Wilson {\it et al.}~\cite{wils00}, both
for the position and magnitude of the hardness peaks. The maxima
(minima) of $\Delta_2E_b(N)$ imply that there are values for which it
is more difficult (easier) to add an atom to the cluster. Moreover, as
N~$\gg 1$ the plot becomes quite smooth, which constitutes an
indication that the cluster can incorporate an additional atom without
major hindrance.

\subsection{Collisions between a single Au atom and a  Au cluster }
\label{subsec:1onN}

Next we report the results of our simulations of the collisions
between a single Au atom and a variable size Au cluster, for several
values both of the initial per atom energy $E$ and of the impact
parameter $b$. The precise details that describe the collision
process are as follows: at time $t=0$ the centers of mass of the atom
and the cluster are placed along the $x$-axis, $\pm10$~{\AA}~ away
from the origin, respectively. The atom is located a distance $b$
away from the $x$-axis, along the $y$-axis on the $xy$ plane. The
principal symmetry axes of the various clusters are aligned
perpendicular to the direction of motion, that is parallel to $y$.
The impact parameter $b$ is varied between 0 (head-on collision) and
7~{\AA}; the latter corresponds to the distance where the interaction
potential effectively vanishes, since the average radius of a cluster
with $12 \leq N \leq 14$ is less than 3~{\AA}~ and the interaction is
cut off at 5.5~{\AA}. The projectile energy is varied between 0.1 and
1.5~eV per atom, in steps of 0.2~eV (8 values in all). The maximum
energy $E=1.5$~eV per atom corresponds to approximately one half of
the cluster binding energy.  After the collision takes place the
resulting fragments are stabilized, by gradual cooling through the
rescaling of the internal velocities. Finally, we analyze the data
characterizing the collision fragments for several special cases. 

We consider two categories: i)~the scattering of a single gold atom
against clusters with N$=12$, $13$ and $14$ atoms, which is dealt with
here; and, ii)~the scattering of a variable size projectile (N$=12$,
$13$ y $14$) on a variable size target with a similar number of Au
atoms, which is presented in \ref{subsec:MonN}.  Throughout we use the
concepts of low and high energies, and small and large impact
parameters. Low energies are defined to be in the range 0.1$\leq E
\leq $0.7~eV and large within 0.7$\leq E \leq $1.5~eV.  Similarly,
small impact parameters cover the range $0 \leq b \leq 3$~{\AA}, and
large is defined as $3 \leq b \leq 7$~{\AA}. The upper bounds on $E$
and $b$, as mentioned above, are related to the binding energy of the
cluster and the distance at which the projectile does not interact
with the target, respectively.

\begin{figure}[!h]
  \centering
  \includegraphics[height=6.3cm]{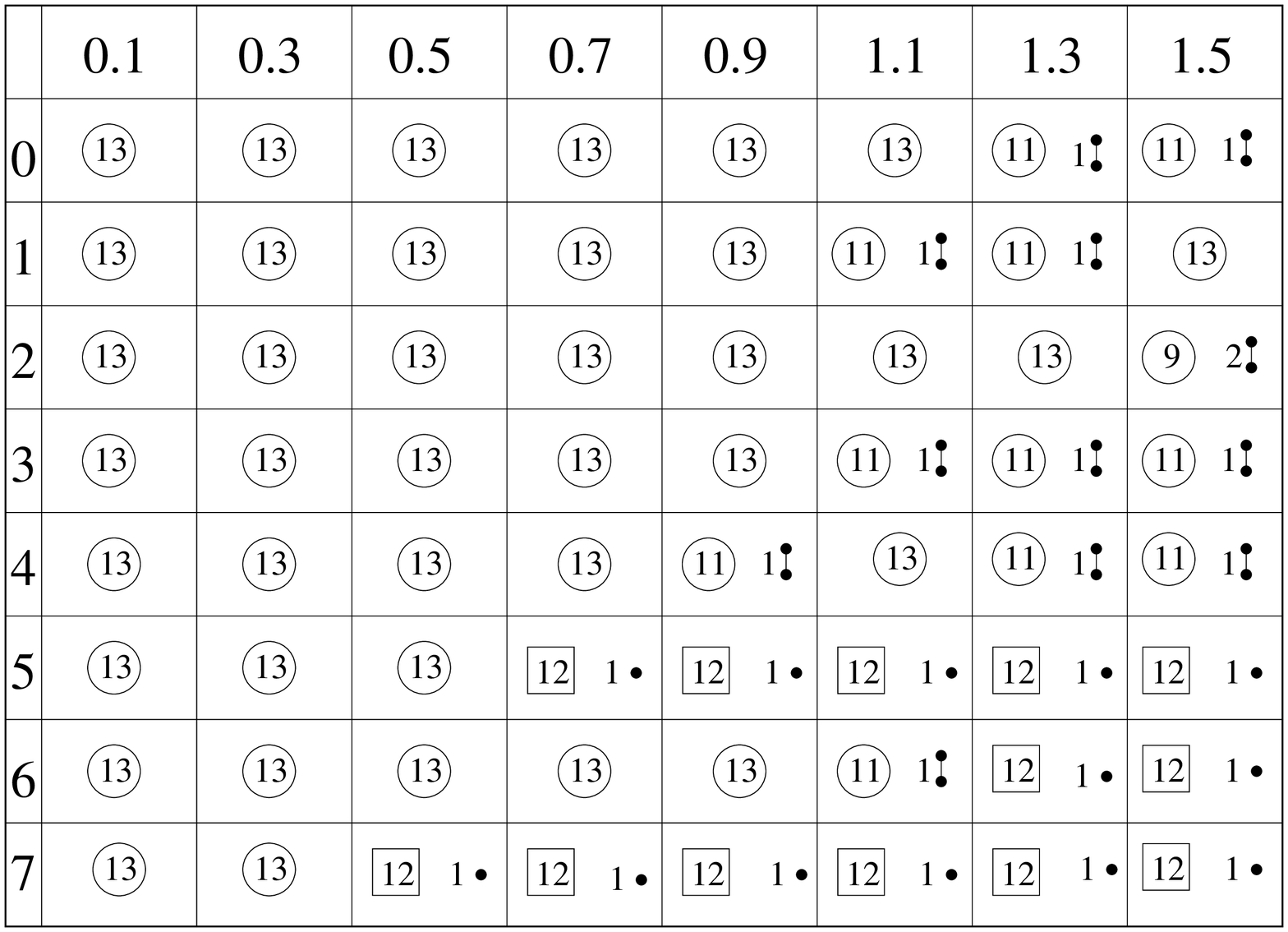}
  \caption{One Au on 12 atom cluster collision. The numbers on the
    left column denote the impact parameter $b$, measured in {\AA}, and
    the top row the average energy per atom $E$ in eV.}
  \label{fig:1on12}
\end{figure}

\noindent {\bf One atom on 12} \ Because of the rich variety of
results, and to facilitate their understanding by the reader, we have
chosen to illustrate them by means of figures. In particular
Fig.~\ref{fig:1on12} describes the one gold atom collision with a 12
atom cluster. In this and ensuing figures the columns correspond to
different values of the energy per atom (in eV), while the rows
correspond to several different impact parameters in {\AA}~ units. Each
entry characterizes a collision on the basis of the following symbols:
single atoms, dimers and trimers are represented by dots, two dots
joined by a line, and three dots that form a triangle, respectively.
When several of these collision fragments are generated we denote
their number by a factor in front of the corresponding symbol. If the
fragment contains four or more atoms the symbolic representation is a
circle. Finally, if the number of atoms in the target is not altered
after the collision, we stress this fact representing it by a square
with the original number of target atoms in its interior, as long as
N$ \geq 4$.

In Fig.~\ref{fig:1on12} it is noticed that for low energies \linebreak
and small impact parameter ($0 \leq b \leq 3$~{\AA}), projectile and
target fuse into a 13 atom cluster. When the energy is increased to
$0.9 \leq E \leq 1.5$~eV, keeping the impact parameter fixed,
coalescence is observed for a few cases, while more often 11 or 9 atom
clusters, and one or two dimers, respectively, are generated. For
larger impact parameters, $4 \leq b \leq 7$, fusion is present in the
low energy region, but for a few cases target and projectile remain
unaltered.  Finally, for large $b$ and $E$ values we observe a few
cases of coalescence, some 11 atom clusters plus a dimer and many instances
(denoted by squares) in which projectile and target size do not
change.

\begin{figure}[!h]
  \centering
  \includegraphics[width=9cm]{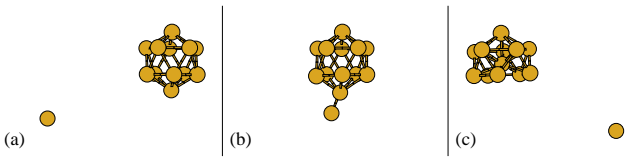}
  \caption{Collision between a single Au atom and a 12 atom
    cluster. The single atom approaches the cluster from the left (a),
    and exchanges energy (b). Finally, in (c) one of the cluster atoms
    is displaced towards the target center, while the projectile
    continues along its trajectory.}
  \label{fig:secuencia}
\end{figure}

A particularly interesting scattering process, illustrated in
Figs.~\ref{fig:secuencia}, occurs for an energy of 4~eV and impact
parameter 1.8~$<b<$~2.2~\AA. When the projectile approaches the target
(Fig.~\ref{fig:secuencia}a) it attracts the nearest (lowest in
Fig.~\ref{fig:secuencia}b) atom, but without removing it from the
cluster. This generates a large energy transfer, which in turn induces
large amplitude vibrations in the ``lowest'' atom, as the projectile
leaves the scene (Fig.~\ref{fig:secuencia}c). As a consequence of
these large amplitude vibrations this particular atom does overcome
the energy barrier and ends up at the center of the cluster
(Fig.~\ref{fig:secuencia}c).  We have estimated a lower bound for this
energy barrier of 0.03~eV. Quite remarkably this lower symmetry
cluster has an energy slightly smaller than the fully symmetric 12
atom cluster we accepted above as the stable configuration.

\noindent In fact, the original 12($I_h$) cluster has a binding energy $E_b =
-3.03$~eV, while the energy of the less symmetric configuration
12(C$_{5v}$) equals $E_b = -3.09$~eV, which is 2\% lower. Thus,
we are faced with the question of why this lower energy configuration
was not obtained in the minimization process we reported above. The
explanation of this apparent contradiction is related to the fact that
the energy barrier the atom has to overcome, to shift to the center of
the cluster, is rather large and cannot be achieved in a minimization
process that starts with an icosahedron and allows only small
displacements from the original equilibrium positions. It is worth
mentioning that this asymmetric structure was also asigned minimum
energy by Wilson and Johnston~\cite{wils00}. However, they argue that
the bond compression of the 12(C$_{5v}$) structure generates a
repulsion strong enough to destabilize it in favor of the icosahedron. 

\noindent {\bf One atom on 13} \ When a Au atom collides with the ``magic
number'' N=13 cluster the results are not much different from the
previous 1 on 12 case.  However, there is a larger number of coalescence
cases, many instances of dimer formation, and also instances are
observed where there is no change in the number of atoms of projectile
and target.  For example, when $E = 0.9$~eV, there is fusion for $b
\leq 2$, dimer plus a 12 atom cluster generation for $b=3$, fusion for
$b=4$, no change $b=5$, and fusion for $b=6$ and $b=7$~{\AA}.

\noindent {\bf One atom on 14} \ In this case the results fall into only two
categories: either target and projectile fuse or they remain
unaltered. Fusion prevails for values of $b\leq 5$~{\AA}. For $b >
5$~{\AA}~ there is coalescence only for very small energies $E \leq 0.3$,
and no change in the number of atoms in projectile and target for
larger energies. This small $b$ large $E$ behavior can be understood
as follows: for low impact parameter the collision gives rise to
violent cluster vibrations and deformations, which precludes the
trapping of the projectile. Instead, larger $b$ values induce less
drastic cluster deformations and sufficient attraction to fuse
projectile and target.

\noindent {\bf One atom on 12: cluster rotation} \ In all preceding
cases the cluster symmetry axis was taken to be parallel to the $y$
axis, that is, perpendicular to the initial projectile velocity. Now
we align the cluster symmetry axis parallel to the $y$ axis. A
qualitative change is observed: single atoms as a result of the
collision. In fact, for low energies and small $b$ there is fusion.
However, for $E>0.7$~eV several alternatives are observed: either
coalescence, or 8, 9 and 11 atom cluster formation accompanied by the
creation of dimers and single atoms. When $b> 3$~{\AA} we obtain 11
atom clusters plus a dimer or a pair of isolated atoms.  Finally, for
the largest $b=7$~{\AA}~ value no atomic reordering is observed.
However, we notice that in general the overall structure of the
results is equivalent to the perpendicularly oriented cluster impact
discussed above, which allows us to concentrate on cluster collisions
without paying much attention to their relative spatial orientation.

\subsection{Collisions between two Au clusters }
\label{subsec:MonN}

Now we turn to the problem of the collision of two clusters of a few,
but in general different, number of atoms. 

\noindent {\bf One dimer on a 12 atom cluster} \ 

Fusion is obtained for low energies ($E\leq 0.7$~eV) and practically
all values of the impact parameter $b$. In many instances the end
result is a dimer and a 12 atom cluster, which are the outcome
after a complex dynamic interaction that, finally, yields a
reconstruction into the two original clusters.  For $E>0.7$~eV and
small $b$ we notice a diversity of results: single atoms plus clusters
of 5, 6, 7, 8, 10 and 11 atoms are obtained, which reflects the fact
that the more complex the projectile the richer the variety of
collision fragments.

\noindent{\bf One 12 atom cluster on another 12 atom cluster } \
Fusion is observed only for low energies ($E=0.1$~eV per atom) over the
whole range of impact parameters $0 \leq b \leq 7$~{\AA}.  For large
energies and small impact parameter collisions, either large
fragments plus a couple of dimers or trimers, or dimers and single
atoms are generated. For large $E$ and large $b$, collisions without
cluster size rearrangement are predominant.

\noindent{\bf One 13 atom cluster on a 12 atom cluster } \ 
Again there is coalescence for $E\sim$~0.1~eV and $0 \leq b \leq 7$~{\AA},
while for large $E$ and large $b$ collisions without rearrangement
predominate. In the $0\sim b \sim $~5~{\AA} and $E\geq 0.5$~eV region
large and medium size fragments plus trimers, dimers and single atoms
are produced. Total break up of the cluster is seen almost exclusively
for small $b$ and large $E$ collisions.

\noindent{\bf One 14 atom cluster on a 12 atom cluster } \ 
Again here the results are quite similar to the previous ones. For
small $b$ there is fusion, while for large $b$ and large $E$ the
collision does not modify the size of the colliding clusters.
Moreover, for a large region of parameter space ($0\leq b \sim 5$~{\AA}~
and $0.5\sim E \sim 1.5$~eV) a whole variety of fragments does result:
large fragments plus dimers, medium size fragments and finally, in for
the largest energies, total cluster breakup into small pieces.

\noindent{\bf One 13 atom cluster on a 13 atom cluster } \ 

\begin{figure}[!h]
  \centering
  \includegraphics[height=6.3cm]{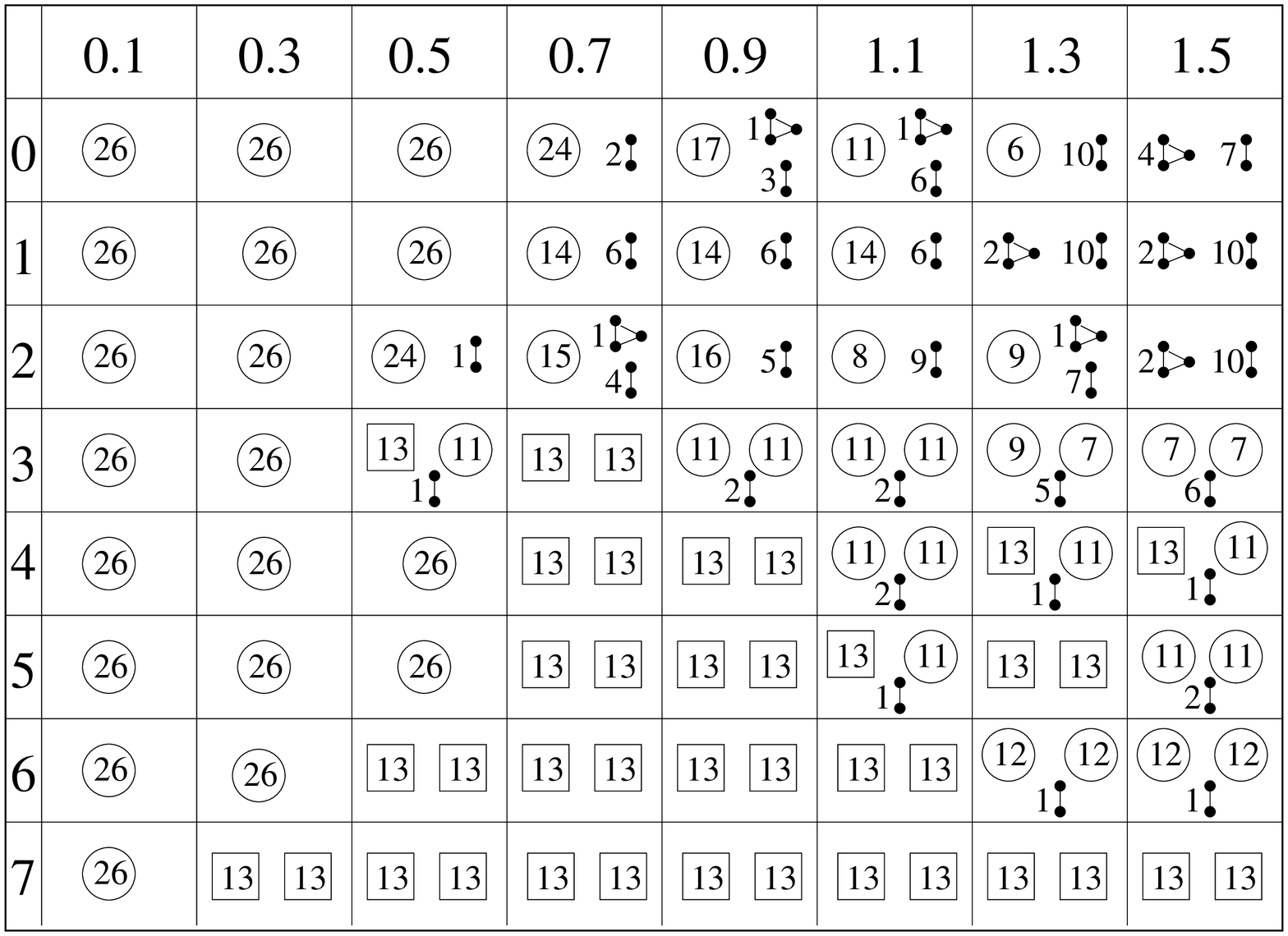}
  \caption{Collision between two 13 atom clusters. The numbers on the
    left column denote the impact parameter $b$, measured in {\AA}, and
    the top row the average energy per atom $E$ in eV.}
  \label{fig:13x13}
\end{figure}

The collision of two ``magic number'' clusters yields the rich variety
of results illustrated in Fig~\ref{fig:13x13}. It is readily noticed
that there are many notable exceptions to the general trends observed
in the preceding cases, and which are only present for this particular
case. This is specially noticeable in the upper right of the figure,
which illustrates the parameter values for which dimers, trimers and
clusters of 6, 7, 8, 9, 11, 12, 14, 15, 16, 17 and 24 atoms are
generated. However, again fusion is observed for low energy collisions
($E\sim 0.1$~eV) for all $b$ values, as well as no rearrangement
collisions for large impact parameters, which are the final outcome of
a complex dynamic interaction that in the end yields a reconstruction
into two clusters equal to the original ones.

\noindent{\bf One 14 atom cluster on a 13 atom cluster } \ 

A low energy coalescence region, as well as a large impact parameter zone
where after the collision projectile and target rearrange into their
original structures, is again obtained. Also, and just as in previous
cases, for large $E$ and small $b$ a diversity of fragments (dimers,
trimers, and 5, 6, 7, 8 and 9 atom clusters) do result. In addition
there are several cases with a single Au atom exchange between
projectile and cluster, such that the original (14, 13) pair is
converted into a (15, 12) one. However, in this same $b$ and $E$
region when three, instead of two, collision fragments are generated,
always a dimer (and not a single atom) is created. This dimer
originates in the 14 atom projectile, yielding in the end a 12 atom
and a 13 atom cluster plus a dimer, as collision fragments.

\section{Conclusions}
\label{sec:concl}

The dynamics of the collision process of gold clusters has been
investigated by means of classical molecular dynamics in combination
with embedded atom (EA) potentials. First, the reliability of the EA
potentials was confirmed by comparison with {\it ab initio} values,
finding that EA is in good agreement for the cluster sizes we
considered, an agreement which improves as N increases. Next,
structural characteristics and the symmetry of the various Au clusters
were obtained and contrasted with published results~\cite{wils00}.

Several type collisions were investigated, finding regions of
coalescence, fragmentation and scattering. Which of these outcomes
actually occurs depends mainly on the values of the projectile energy
$E$ and the impact parameter $b$. Coalescence is dominant for low
energies ($E < 0.7$~eV) and small impact parameters ($b<3$~\AA).
Simple scattering, with no change in the size and structure of the
colliding clusters, prevails for large $E$ and $b$ values. For large
energies and small impact parameters fragmentation and scattering are
generally the case. For large $E$ and large $b$ scattering is the most
probable outcome. When the cluster does breaks up the main collision
products, apart from large fragments, are dimers. On the other hand
cluster coalescence provides a viable mechanism to generate larger
cluster sizes. It is also of interest that the collisions themselves
turn out to be rather insensitive to the relative orientation of the
projectile and target main symmetry axes, and that cluster collisions
can generate metastable structures, which are usually not accesible
due to the existence of a potential barrier.

%
%

%
%


\newpage 
\onecolumn

\begin{center}
\begin{tabular}{|lccccc|}
\hline
{\bf Cluster} &{\bf Method} &{\bf Symmetry} & $E_b$~[eV] & $R$~[{\AA}] &
$v$~[\AA $^3$]\\ \hline \hline
\multirow{3}{20mm}{3 atoms}
& EA & C$_{2v}$& 2.405  & 1.937--2.438 &  3.317\\ 
& CP & C$_{2v}$& 2.182  & 2.607--2.753 &  4.225\\ 
&WJ&D$_{3h}$ & 1.759  & 2.780 & 4.133 \\ \hline
\multirow{5}{20mm}{4 atoms}
& EA & D$_{2d}$& 2.313  & 1.904--2.874 &  3.996\\ 
& CP (a) &D$_{2d}$ & 2.234  & 2.570--3.120 & 5.857\\ 
& CP (b) &D$_{2h}$ (planar) & 2.517 & 2.700 &  2.346\\ 
& CP (c) &C$_{2v}$ (planar) & 2.504 & 2.648 &  29.759\\ 
&WJ&T$_{d}$ &2.178 & 2.793 & 5.004\\ \hline
\multirow{4}{20mm}{5 atoms}
& EA & O$_{h}$& 2.685  & 2.507 & 7.287 \\ 
& CP (a)& O$_{h}$& 2.464 & 2.801 & 4.672\\ 
& CP (b)&C$_{2v}$ (planar)  & 2.652 & 2.715  & 7.706 \\ 
 &WJ      &O$_{h}$ &2.382 & 2.801 &  4.290 \\ \hline
\multirow{4}{20mm}{6 atoms}
& EA & O$_{h}$& 2.825  & 2.536 & 5.767 \\ 
& CP (a)& O$_{h}$& 2.523  & 2.834 & 8.663 \\ 
& CP (b)& D$_{3h}$ (planar) & 2.870  & 2.712 &  4.365 \\ 
&WJ&O$_{h}$ &2.574 & 2.789 & 7.667 \\ \hline
\multirow{3}{20mm}{8 atoms}
& EA & D$_{2d}$& 2.937  & 2.577 &  5.512\\ 
& CP & D$_{2d}$& 2.857 & 2.837 & 7.385\\ 
& WJ & D$_{2d}$ & 2.736 & 2.790 & 6.792 \\ \hline
\multirow{3}{20mm}{10 atoms}
& EA & D$_{4d}$& 2.999  & 2.594 &  9.794\\ 
& CP & D$_{4d}$& 2.914 & 2.829 & 36.225\\ 
& WJ & D$_{4d}$ & 2.837 & 2.778 & 30.854  \\ \hline
\multirow{5}{20mm}{12 atoms}
& EA (a)& I$_{h}$  & 3.027  & 2.604 &  15.190\\ 
& EA (b)&C$_{5v}$& 3.089 &  2.693 & 21.142\\ 
& CP (a)& C$_{5v}$& 2.879  & 2.896 & 33.032\\
& CP (b)& D$_{4h}$ & 2.962  & 2.814 & 8.780\\  
&WJ&I$_{h}$ &2.886 & 2.776 &   18.395    \\ \hline
\end{tabular}
\label{table:binding_E}
\end{center}

\end{document}